\documentclass[a4paper,11pt]{article}
\usepackage{pos}
\usepackage{physics}
\usepackage{wrapfig,bm}

\newcommand{\bvec}[1]{\boldsymbol{#1}}

\title{Study of 3-dimensional SU(2) gauge theory with adjoint Higgs as a model for cuprate superconductors}
\ShortTitle{3-dimensional SU(2) gauge theory with adjoint Higgs}

\author[a]{Guilherme Catumba}
\author[b]{Atsuki Hiraguchi}
\author[c]{George W.-S. Hou}
\author[d]{Karl Jansen}
\author[c,e]{Ying-Jer Kao}
\author[f,g,h]{C.-J. David Lin}
\author[a]{Alberto Ramos}
\author*[c,i]{Mugdha Sarkar}

\affiliation[a]{Instituto de Física Corpuscular (IFIC), CSIC-Universitat de Valencia.\\
46071, Valencia, Spain}

\affiliation[b]{CCSE, Japan Atomic Energy Agency, 178-4-4, Wakashiba, Kashiwa, Chiba 277-0871, Japan}

\affiliation[c]{Department of Physics, National Taiwan University,\\
Taipei 10617, Taiwan}

\affiliation[d]{Deutsches Elektronen-Synchrotron DESY,\\
Platanenallee 6, 15738 Zeuthen, Germany}

\affiliation[e]{Center for Theoretical Physics and Center for Quantum Science and Technology, National Taiwan University,\\
Taipei, 10607, Taiwan}

\affiliation[f]{Institute of Physics, National Yang Ming Chiao Tung University,\\ 
1001 Ta-Hsueh Road, Hsinchu 30010, Taiwan}

\affiliation[g]{Center for High Energy Physics, Chung-Yuan Christian University,\\
Chung-Li 32023
, Taiwan}

\affiliation[h]{Centre for Theoretical and Computational Physics, National Yang Ming Chiao Tung University,\\ 
1001 Ta-Hsueh Road, Hsinchu 30010, Taiwan}

\affiliation[i]{Physics Division, National Centre for Theoretical Sciences,\\ 
Taipei 10617, Taiwan}

\emailAdd{mugdha.sarkar@gmail.com}

\abstract{We study a 3-dimensional SU(2) gauge theory with 4 Higgs fields which 
transform under the adjoint representation of the gauge group, 
that has been recently proposed by Sachdev et al. to explain the physics 
of cuprate superconductors near optimal doping. The symmetric 
confining phase of the theory corresponds to the usual 
Fermi-liquid phase while the broken (Higgs) phase is associated 
with the interesting pseudogap phase of cuprates. We employ the
Hybrid Monte-Carlo algorithm to study the phase diagram of the theory. 
We find the existence of a variety of broken phases in qualitative accordance 
with earlier mean-field predictions and discuss their role in cuprates.
In addition, we investigate the behavior of Polyakov loop to probe the 
confinement/deconfinement phase transition, and find that the Higgs phase 
hosts a stable deconfining phase consistent with previous studies.}

\FullConference{The 40th International Symposium on Lattice Field Theory (Lattice 2023)\\
July 31st - August 4th, 2023\\
Fermi National Accelerator Laboratory\\}

\begin{document}
\maketitle

\section{Introduction}
\begin{wrapfigure}{r}{0.4\textwidth}
  \begin{center}
    \includegraphics[width=.35\textwidth,,trim={0 0 18cm 0},clip]{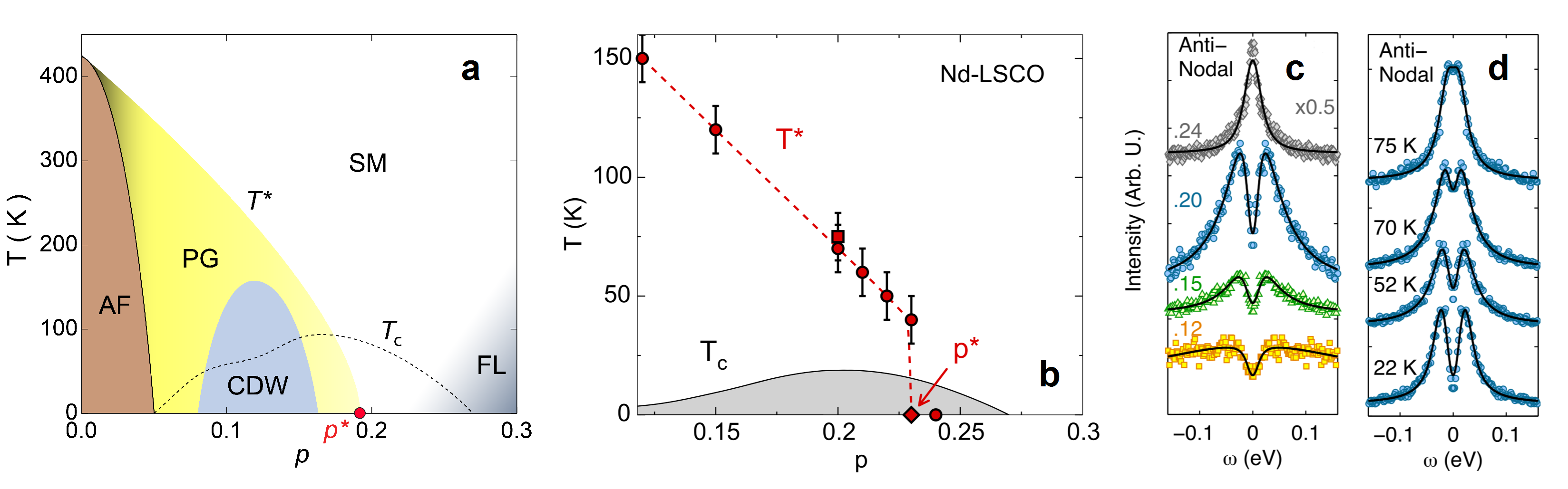}
  \end{center}
  \caption{Schematic phase diagram of cuprates in the temperature ($T$) and hole-doping density ($p$) directions, reproduced from Ref.~\cite{CMPreview}.}
  \label{fig:schematicpd}
\end{wrapfigure}

A complete understanding of the phase diagram of cuprate superconductors has been a long-standing problem in the condensed matter community since their Nobel prize-winning discovery in the 1980s (for a review, see for example, \cite{CMPreview}). In Fig.~\ref{fig:schematicpd}, we show a schematic phase diagram of cuprates as a function of the temperature $T$ and hole-doping density $p$.
At zero and low doping density, the behavior of cuprates as a Mott insulator with antiferromagnetic (AF) order is well understood in terms of the high repulsive interactions among the electrons at every $Cu$ site in the $CuO_2$ planes. 
At high $p>0.3$, there exists a familiar metallic Fermi liquid (FL) phase. The superconducting phase (SC) exists at intermediate doping upto temperatures $T_c$ as high as $150K$ as depicted by the region below the dotted line. At densities above an optimal doping $p^*$, the SC phase is consistent with the BCS-BdG theory of superconductivity. What is still not well understood is the unconventional SC state which coexists along with a peculiar psuedogap phase (PG) at intermediate doping below $p^*$. At high $T$ and $p$, there exists a yet another puzzling strange metal phase (SM) which will not be the subject of discussion in this work. 
In this work we are interested in the ground state near optimal doping, where experimental studies have revealed a plethora of interesting phenomena including charge density wave (CDW) and nematic orders, time-reversal symmetry breaking that are revealed from beneath the SC phase at strong magnetic field \cite{CMPreview, fujitascience, yangscience, girodnature, xiaprl}. 
An understanding of the microscopic mechanism of the pseudogap phase and its underlying states is thus important to understand the origin of the unconventional superconductivity in cuprates. 

\section{Proposal}
In Ref.~\cite{Sachdev:2018nbk}, Sachdev \textit{et al} proposed a $(2+1)$-dimensional $SU(2)$ gauge theory of the fluctuations of the incommensurate spin-density-wave (SDW) order to explain various phenomena of the pseudogap phase of cuprates around optimal doping. The key idea is to transform the SDW order parameter $\bm{S}$ to a spacetime-dependent rotating reference frame in spin space,
\begin{equation}
\bm{\sigma} \cdot \bm{S}_i = R_i \bm{\sigma} R_i^\dagger \cdot \bm{H}_i,
\end{equation} 
where $\bm{\sigma}$ are the Pauli matrices, subscript $i$ denotes the site index in $2D$ space, $R_i$  is a spacetime-dependent $SU(2)$ group element and $\bm{H}_i$ is the rotated spin magnetic moment. This transformation induces a local $SU(2)$ gauge invariance and the emergent $R_i$ and $\bm{H}_i$ fields transform as : $R_i \to R_i V_i^\dagger, \bm{\sigma} \cdot \bm{H}_i \to V_i \bm{\sigma}V_i^\dagger \cdot \bm{H}_i$, where $V_i \in SU(2)$. The $\bm{H}_i$ field transforms in the adjoint representation of $SU(2)$ and behaves as an adjoint Higgs field. 

In the intermediate doping region of hole-doped cuprates, experiments indicate existence of an SDW order which is collinearly polarized at incommensurate $2D$ wave vectors $\bm{K}_x$ and $\bm{K}_y$. To capture this spatial ordering, the rotated order parameter $\bm{H}_i$ can be parametrized as
\begin{equation}
\bm{H}_i = \mathrm{Re}\left[\bm{\mathcal{H}}_x e^{i\bm{K}_x\cdot \bm{r}_i} + \bm{\mathcal{H}}_y e^{i\bm{K}_y\cdot \bm{r}_i}\right], \label{Eq:spatial}
\end{equation}
where $\mathcal{H}_x$ and $\mathcal{H}_y$ are complex fields which also transform under the adjoint representation of $SU(2)$. In terms of the emergent fields, one constructs an $SU(2)$ gauge theory Lagrangian coupled to $N_h=4$ real adjoint Higgs fields as given in Eq.~2.1 in Ref.~\cite{Sachdev:2018nbk}. The Higgs potential is constrained by the global symmetries of the system. The Higgs field is further coupled to fermionic degrees of freedom, carrying electric charge, which are not considered in this study.
The usual symmetric (confinement) phase of the gauge-Higgs theory corresponds to the FL phase at high $p$. On the other hand, the broken (Higgs) phase with a Higgs condensate, coupled to the fermionic degrees of freedom, induces ordering that can be mapped to the various orders within the pseudogap phase. Under appropriate gauge-fixing, the $SU(2)$ gauge symmetry can be broken into either $U(1)$ or $Z(2)$ for $N_h=4$ \cite{Sachdev:2018nbk,Scammell:2019erm,Bonati:2021tvg}, in contrast to the $N_h=1$ case\footnote{The $N_h=1$ theory can be mapped to the case of electron-doped cuprates \cite{Sachdev:2018nbk}.} \cite{Nadkarni:1989na,Hart:1996ac}.

\section{Lattice implementation}
We discretize the above-mentioned $SU(2)$ gauge theory on an Euclidean lattice to study the phase diagram. The lattice action for $3$-dimensional $SU(2)$ gauge theory with $N_h=4$ adjoint Higgs fields can be written as
\begin{gather}
S = \beta \sum_x \sum_{\mu < \nu}^3 \left(1 - \frac{1}{2} \Tr U_{\mu\nu}(x) \right) 
   - 2{\kappa} \sum_{x,\mu} \sum_{n=1}^4 \Tr \left( \Phi_n(x) U_\mu(x)\Phi_n(x+\hat\mu) U^\dagger_\mu(x)\right) 
   + V(\{\Phi_n\}) \, ,
   \label{Eq:genaction}
\end{gather}
where the first term is the usual gauge Wilson action composed of plaquettes $U_{\mu\nu}(x)$ and the second term describes the coupling of the Higgs fields $\Phi_n^i(x)$ to the gauge field $U_\mu(x)$. We rewrite the complex Higgs fields $\mathcal{H}_{x/y}$ in Eq.~\ref{Eq:spatial} as 
  $\bvec{\mathcal{H}}_{x} = \varphi_1 + i\varphi_2$,
  $\bvec{\mathcal{H}}_{y} = \varphi_3 + i\varphi_4$,
with real fields $\varphi_n^i(x)$ ($n=1,\ldots,4$ and $i=1,2,3$). The Higgs fields appearing in Eq.~\ref{Eq:genaction} are given in terms of the $\varphi_n^i(x)$ as $\Phi_n^i = \sqrt{a/\kappa} \varphi_n^i$ , with $a$ being the lattice spacing. Constrained by the global symmetries of the cuprate system, \textit{i.e.}, time-reversal and square-lattice symmetries, the potential is given as
  \begin{align}
  V(\{\Phi_n\}) = 2 \sum_x &\sum_{n=1}^4 V_{nn} + \lambda \sum_x \sum_{n=1}^4 \left( 2 V_{nn} - 1\right)^2 \nonumber \\ 
  + 4\lambda \sum_x&\sum_{n \neq m}^4 V_{mm}V_{nn} 
  + \hat{u_1}\sum_x \sum_{m,n}^4 g_{mn} V_{mm}V_{nn}
  + 4\hat{u_3}\sum_x \sum_{n \neq m}^4 V_{mn}V_{mn} \nonumber \\
  + \hat{u_2}\sum_x&  \left[\sum_{n=1}^4 (V_{nn})^2 
  + 4(V_{12})^2 + 4(V_{34})^2 - 2V_{11}V_{22}
  - 2V_{33}V_{44} \right]  \,, 
  \end{align}
using the notation $V_{mn}=\Tr(\Phi_m\Phi_n)$. The Higgs fields are expressed as $\Phi_n=\Phi_n^i T^i$, where $T_i = \sigma_i/2$ ($\sigma_i$ are Pauli matrices), such that
$2\Tr(\Phi_m\Phi_n)=\sum_{i=1}^{3}\Phi_m^i\Phi_n^i$. The $g_{mn}$ in the 
4\textsuperscript{th} term are given by $g_{mm}=g_{12}=g_{34}=1$ and -1 otherwise. 
The dimensionful couplings in the continuum theory (Eq.~(2.1c) of \cite{Sachdev:2018nbk}) are related to the dimensionless lattice couplings as follows
  \begin{gather} \label{Eq:cont_lat_relations_nh4}
  \beta = \frac{4}{g^2 a}, \quad \lambda=u_0 a \kappa^2, 
  \quad s = \frac{1}{a^2}\left[\frac{1}{\kappa} - 3 - 
  \frac{2\lambda}{\kappa}\right], \quad
  \hat{u_i} = u_i a \kappa^2 \quad (i=1,2,3).
  \end{gather}

\begin{figure}
  \centering
   \includegraphics[width=0.39\textwidth]{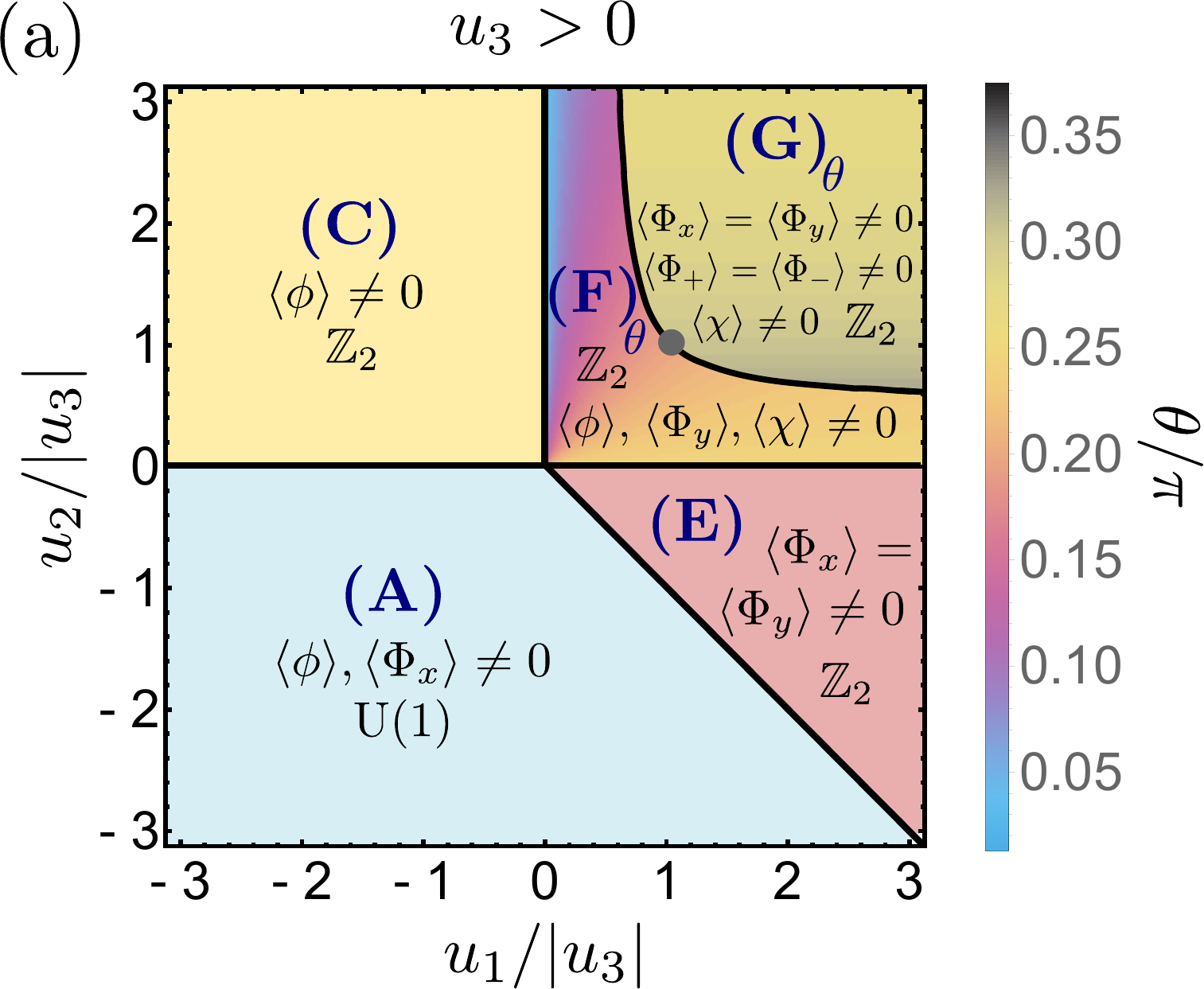}
   \includegraphics[width=0.2925\textwidth]{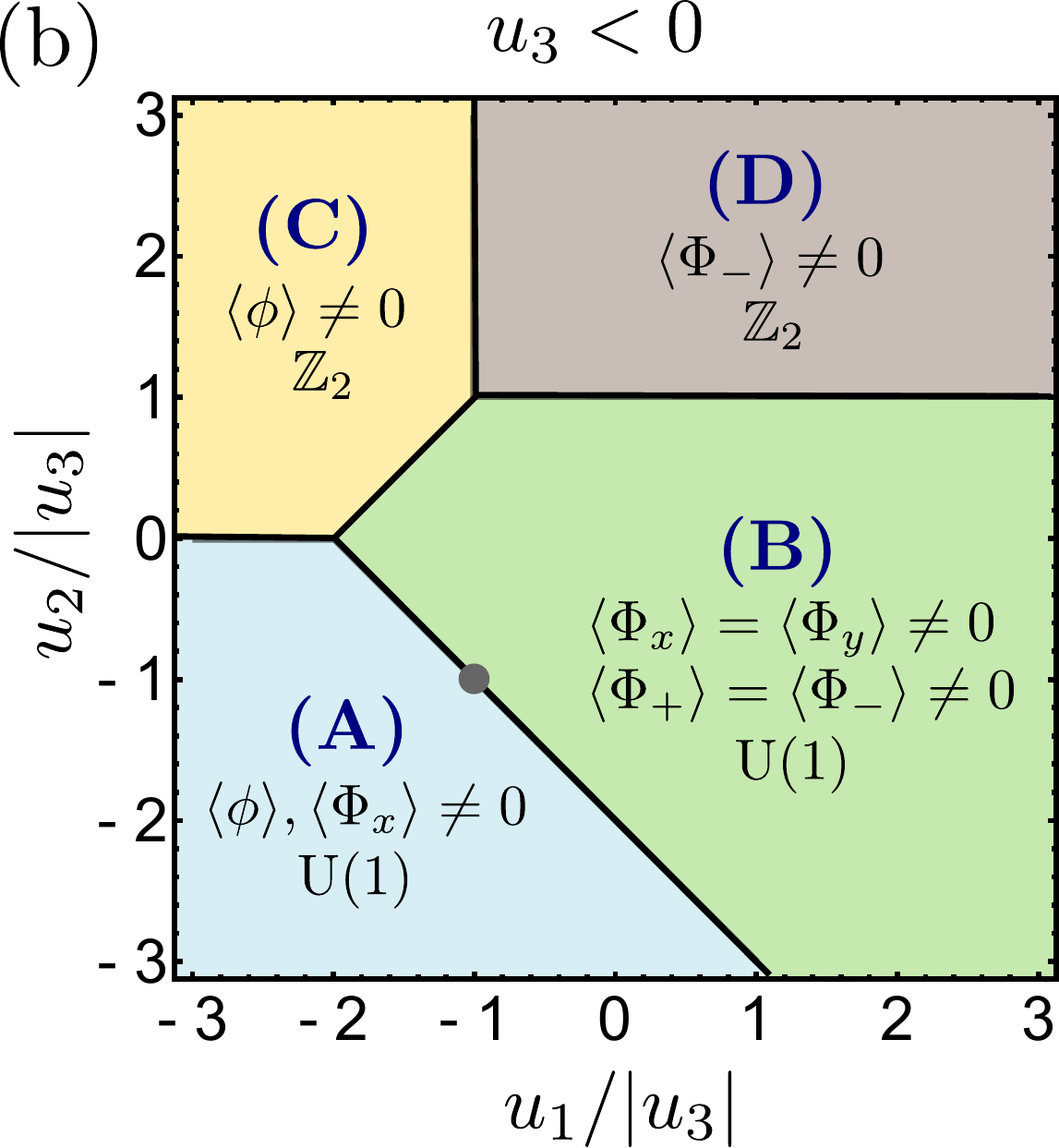}
	\caption{{Phase diagram of the Higgs region as a function of the quartic couplings $u_i$ at fixed $\kappa$ and $\lambda$, obtained from a mean-field analysis without gauge fields in  Ref.~\cite{Sachdev:2018nbk}.}
	} \label{fig:meanfld}
 \end{figure}
The Higgs phase in this theory consists of several types of broken phases, labelled (A)-(G), which are obtained upon breaking of different continuous and discrete global symmetries of the Higgs potential. The Higgs phase diagrams as a function of the quartic couplings $u_i$ in Fig.~\ref{fig:meanfld} have been obtained by a mean-field analysis in \cite{Sachdev:2018nbk}. The Higgs condensate in phases (A) and (B) leads to a $SU(2)\to U(1)$  symmetry breaking pattern while that in phases (C)-(G) leads to the pattern $SU(2)\to Z(2)$. The ordered phases are characterised by the following local gauge-invariant Higgs bilinears 
\cite{Sachdev:2018nbk},
\begin{align}
\phi &= \Phi_1^i\Phi_1^i + \Phi_2^i\Phi_2^i - \Phi_3^i\Phi_3^i - \Phi_4^i\Phi_4^i,\nonumber \\
\Phi_x &= \Phi_1^i\Phi_1^i - \Phi_2^i\Phi_2^i + 2i\Phi_1^i\Phi_2^i, \nonumber \\
\Phi_y &= \Phi_3^i\Phi_3^i - \Phi_4^i\Phi_4^i + 2i\Phi_3^i\Phi_4^i, \nonumber \\
\Phi_+ &= \Phi_1^i\Phi_3^i - \Phi_2^i\Phi_4^i + i(\Phi_1^i\Phi_4^i + \Phi_2^i\Phi_3^i), \nonumber \\
\Phi_- &= \Phi_1^i\Phi_3^i + \Phi_2^i\Phi_4^i + i(\Phi_1^i\Phi_4^i - \Phi_2^i\Phi_3^i)
	\label{Eq:bilinear}
\end{align}
which signal breaking of different square-lattice symmetries leading to nematic order or unidirectional, bidirectional and spiral CDW orders, which have been observed in experiments on hole-doped cuprates. A trilinear gauge-invariant observable, 
\begin{align}
    \chi_{xyy} &= -2\epsilon_{ijk}\left( \Phi_2^i\Phi_3^j\Phi_4^k - i \Phi_1^i\Phi_3^j\Phi_4^k\right)\,, \nonumber \\
    \chi_{yxx} &= -2\epsilon_{ijk}\left( \Phi_4^i\Phi_1^j\Phi_2^k - i \Phi_3^i\Phi_1^j\Phi_2^k\right), \label{Eq:chi}
\end{align}
where $\epsilon_{ijk}$ is the Levi-Civita symbol in the adjoint indices, 
serves as the order parameter for time-reversal symmetry breaking in the cuprate system \cite{scheurer}. The phases (F) and (G), which display time-reversal symmetry breaking alongwith breaking of certain square-lattice symmetries, are phenomenologically appealing possibilities. It should be noted that the observables defined above can have either non-zero real or imaginary parts due to a degeneracy of the broken ground states. Therefore, we compute the expectation value of the absolute value of the complex observables in th absence of any magnetic-field-like terms which lift the degeneracy.

\begin{figure}
  \centering
  \includegraphics[width=0.37\textwidth]{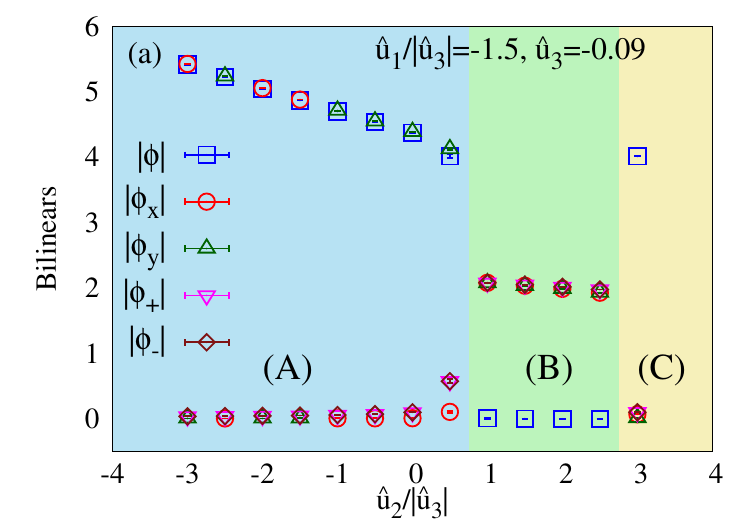}
   \includegraphics[width=0.37\textwidth]{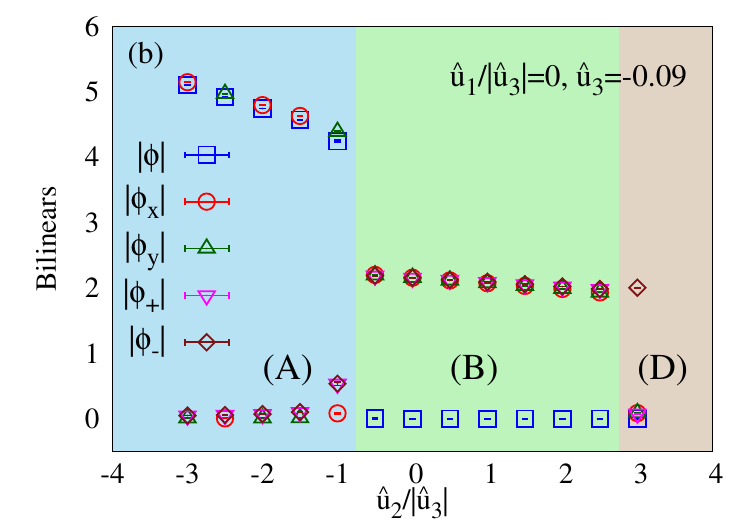}
   \includegraphics[width=0.37\textwidth]{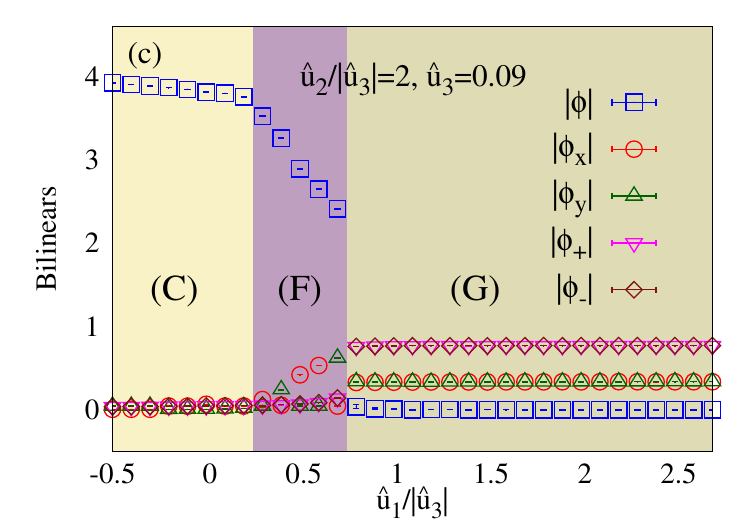}
   \includegraphics[width=0.37\textwidth]{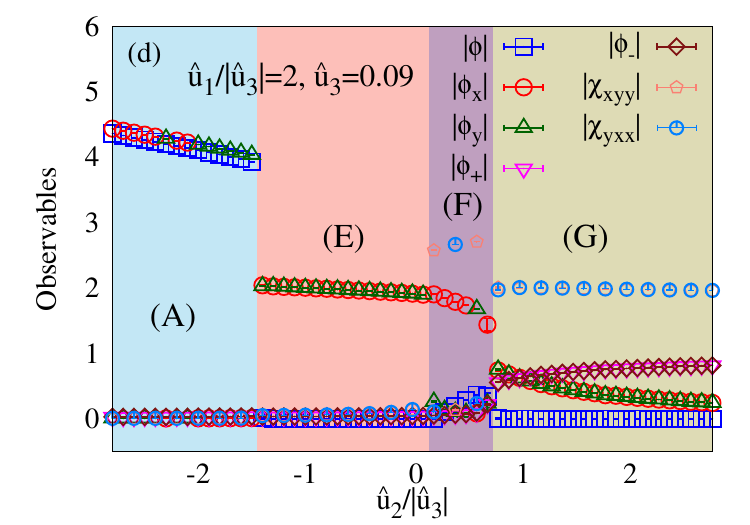}  
  \caption{Expectation values of observables as functions of coupling ratios $\hat{u}_1/|\hat{u}_3|$ or $\hat{u}_2/|\hat{u}_3|$ at fixed $\hat{u}_3$ values. 
	The results have been obtained from simulations on $12^3$ and $16^3$ lattices.}\label{fig:posneg_u3}
 \end{figure} 
 
In \cite{Sachdev:2018nbk}, the possibility of a deconfined phase within the Higgs phase was discussed where $SU(2)$ color-neutral electron-like particles may deconfine into spinon and chargon degrees of freedom having $SU(2)$ charges. 
We, therefore, study the Polyakov loop, which is a true order parameter for confinement-deconfinement transition in a gauge theory with adjoint Higgs, defined as, $L=\frac{1}{N_s^2} \sum_{\vec{x}}  \Tr \prod_{t=1}^{N_t} U_t(\vec{x},t)\,$ on a $N_s^2\times N_t$ lattice.  
 
\subsection{Simulation details}
We perform simulations of the $SU(2)$ gauge theory with $N_h=4$ adjoint Higgs using the Hybrid Monte Carlo algorithm. We have calculated the observables on $3D$ lattices of size $12^3$, $16^3$ and $24^3$. All of the results shown here have been performed at a fixed gauge coupling $\beta=8.0$. 
 
\section{Results}
In \cite{Scammell:2019erm}, a numerical study of the strong gauge coupling limit of the model was performed  with a simplified global $O(N_h=4)$-invariant Higgs potential obtained by setting the couplings $u_1=u_2=u_3$. The existence of two different broken phases, symmetric under $U(1)$ and $Z(2)$, was found in this approximation. Another recent study \cite{Bonati:2021tvg} also found the two symmetry breaking patterns in the $O(4)$ limit but with finite values of the gauge coupling. In our study, we perform numerical simulations with the complete action, proposed in 
\cite{Sachdev:2018nbk}.

Our initial lattice simulations of the theory suggest qualitative agreement with the mean-field phase diagram. We conducted both horizontal and vertical scans of the positive/negative $u_3$ phase diagram as shown in Fig.~\ref{fig:meanfld}. For our scans, we choose the lattice quartic coupling $\hat{u_3}$ to be either $0.09$ or $-0.09$, and tune $\hat{u}_{1}$ ($\hat{u}_{2}$) at fixed $\hat{u}_{2}$ ($\hat{u}_{3}$) such that the ratio $\hat{u_1}/|\hat{u_3}|$ ($\hat{u_2}/|\hat{u_3}|$) varies in the range $[-3.5:3.5]$ (It can be easily seen from Eq.~\ref{Eq:cont_lat_relations_nh4} that any ratio of lattice quartic couplings is equal to its continuum counterpart at a fixed $\kappa$). To ensure that we remain in the broken (Higgs) phase for these range of couplings, the lattice couplings $\kappa$ and $\lambda$ are set to $3.0$ and $1.0$ respectively. We plot the variation of the expectation values of the observables as a function of the quartic couplings in Fig.~\ref{fig:posneg_u3}. For example, in Fig.~\ref{fig:posneg_u3} (d), we show the absolute values of the bilinears as a function of $\hat{u}_2/|\hat{u}_3|$ at a fixed $\hat{u}_1/|\hat{u}_3|=2$ and $\hat{u}_3>0$ (vertical line in the phase diagram Fig.~\ref{fig:meanfld} (a)), which describe the transitions among phases (A)-(E), (E)-(F) and (F)-(G), being in good agreement with the mean-field prediction. The time-reversal symmetry breaking order parameters $\chi_{xyy}$ and $\chi_{yxx}$ are vanishing in all phases except (F) and (G), with either one of them being non-zero in phase (F) and both being non-zero and equal to each other in (G).

\begin{figure}
  \centering
  \includegraphics[width=0.4\textwidth]{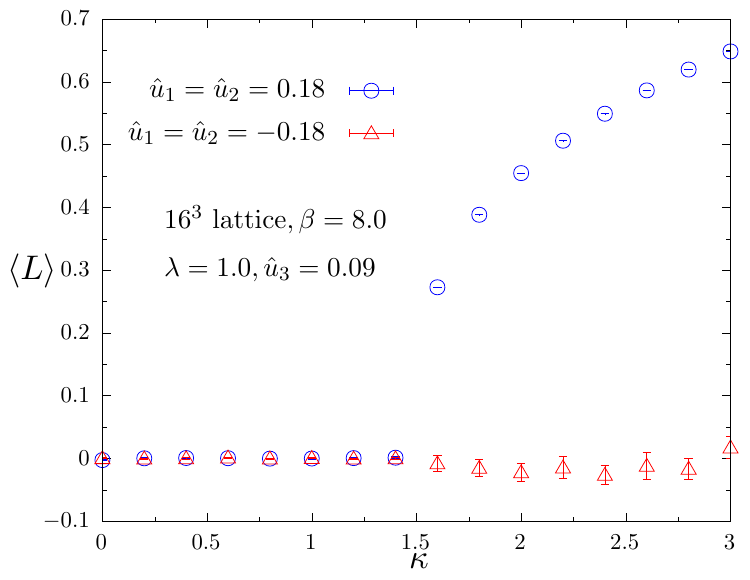}
  \caption{The Polyakov loop expectation value as a function of $\kappa$ for two different sets of fixed values for the quartic couplings $\lambda$ and $\hat{u}_i$. For all positive $u_i$ the broken phase corresponds to phase (G) while for the latter case it corresponds to phase (A).}\label{fig:ploop}
 \end{figure}
 
Next we discuss the Polyakov loop as an order parameter for deconfinement in different phases. In Fig.~\ref{fig:ploop}, we show the expectation value $\expval{L}$ as a function of $\kappa$ across the transition from symmetric to Higgs phases. The symmetric phase of the theory is confining as expected. Interestingly in the broken phase corresponding to symmetry-breaking pattern $SU(2)\to Z(2)$, we find a deconfined phase with $\expval{L}\neq 0$. On the other hand, the broken phase with remnant $U(1)$ symmetry has $\expval{L}$ consistent with zero. In the latter  case, we find that the Polyakov loop fluctuates between the two degenerate minima resulting from the spontaneously broken $Z(2)$ center symmetry group. As a result, the broken phase with remnant $U(1)$ is not a stable deconfining phase. Our findings are consistent with the arguments in \cite{Sachdev:2018nbk} which support a stable deconfining phase only in the phase with remnant local $Z(2)$ symmetry. 

\section{Conclusions and outlook}
We have studied the phase diagram of a $3D$ $SU(2)$ gauge theory with $N_h=4$ adjoint Higgs fields, which has been proposed to describe the pseudogap phase in hole-doped cuprate superconductors near optimal doping \cite{Sachdev:2018nbk}. Our lattice study of the complete action with dynamical gauge fields qualitatively confirms the mean-field prediction of the theory and we demonstrate the presence of distinct broken phases, as indicated by the expectation values of gauge-invariant observables of the Higgs fields. We study the Polyakov loop as an order parameter for deconfinement and find that the broken phase with remnant $Z(2)$ symmetry is a stable deconfining phase whereas the broken phase with remnant $U(1)$ is not.

\section{Acknowledgements}
The work of AH is supported by the Taiwanese
NSTC grant 110-2811-M-A49-501-MY2 and 111-2639-M-002-004-ASP.
The work of CJDL is supported by the Taiwanese
NSTC grant 112-2112-M-A49-021-MY3 and 111-2639-M-002-004-ASP.
AR and GT acknowledge support from the Generalitat Valenciana
(genT program CIDEGENT/2019/040) and the Ministerio de Ciencia e
Innovacion  PID2020-113644GB-I00. 
The work of MS is supported by the Taiwanese
MoST grant NSTC 110-2811-M-002-582-MY2, NSTC 108-2112-M-002-020-MY3 and 111-2124-M-002-013-.
The authors gratefully acknowledge
the computer resources at Artemisa, funded by the European Union ERDF
and Comunitat Valenciana as well as the technical support provided by
the Instituto de Física Corpuscular, IFIC (CSIC-UV), and the 
computer resources at AS, NTHU, NTU, NYCU and DESY, Zeuthen.

\end{document}